\documentclass[12pt]{article}
\usepackage[dvips]{epsfig}
\psfigdriver{dvips}

\begin{document}

%
\def\papertitlepage{\baselineskip 3.5ex \thispagestyle{empty}}
\def\preprinumber#1#2{\hfill \begin{minipage}{4.2cm}  #1
                 \par\noindent #2 \end{minipage}}
\renewcommand{\thefootnote}{\fnsymbol{footnote}}
\newcommand{\beq}{\begin{equation}}
\newcommand{\eeq}{\end{equation}}
\newcommand{\beqa}{\begin{eqnarray}}
\newcommand{\eeqa}{\end{eqnarray}}
\catcode`\@=11
\@addtoreset{equation}{section}
\def\theequation{\thesection.\arabic{equation}}
\catcode`@=12
\relax
\newcommand{\Lim}{\lim_{\Lambda_0\rightarrow -\infty}}

%
%
\papertitlepage
\setcounter{page}{0}
\preprinumber{KEK-TH-987}{hep-th/0411049}
\baselineskip 0.8cm
\vspace*{2.0cm}
\begin{center}
{\large\bf On Unitary/Hermitian Duality in Matrix Models}
\end{center}
\vskip 4ex
\baselineskip 1.0cm
\begin{center}
           {Shun'ya~ Mizoguchi\footnote[1]{\tt mizoguch@post.kek.jp} } 
\\
    \vskip -1ex
       {\it High Energy Accelerator Research Organization (KEK)} \\
    \vskip -2ex
       {\it Tsukuba, Ibaraki 305-0801, Japan} \\
\end{center}
\vskip 10ex
%
\baselineskip=3.5ex
\begin{center} {\large\bf Abstract} \end{center}
Unitary 1-matrix models are shown to be exactly equivalent to 
hermitian 1-matrix models coupled to $2N$ vectors with appropriate 
potentials, to all orders in the $1/N$ expansion. This fact allows us to 
use all the techniques developed and results obtained in hermitian 
1-matrix models to investigate unitary as well as other 1-matrix models 
with the Haar measure on the unitary group. We demonstrate the 
use of this duality in various examples, including: (1) an explicit 
confirmation that the unitary matrix formulation of the ${\cal N}=2$ 
pure $SU(2)$ gauge theory correctly reproduces the genus-1 topological 
string amplitude (2) derivations of the special geometry 
relations in unitary as well as the Chern-Simons matrix models. 
 \vskip 2ex
\vspace*{\fill}
\noindent
November 2004
\newpage
\renewcommand{\thefootnote}{\arabic{footnote}}
\setcounter{footnote}{0}
\setcounter{section}{0}
\baselineskip = 0.6cm
\pagestyle{plain}

\section{Introduction}
Random matrix models \cite{Mehta} have served as cornerstones
for various areas of theoretical physics. 
Hermitian matrix models, 
particularly those in double-scaling limits, were intensively 
studied in the early 90's in connection to two-dimensional quantum gravity 
\cite{2dgravity}. Unitary models, on the other hand, are known to arise as a 
one-plaquette lattice model of two-dimensional Yang-Mills theory \cite{GW}. 
More recently, matrix models, both hermitian and unitary ones, 
have proven to be powerful tools for studying supersymmetric gauge theory 
and topological string theory \cite{DV,DV2}.

Both hermitian and unitary 1-matrix models are reduced, 
after the gauge fixing, to eigenvalue models with different Jacobians.
Despite the difference, they induce at short distances the same interaction  
between the eigenvalues, and share many common properties.
For instance, the critical behaviors 
of both unitary and (even-potential) hermitian 1-matrix models are 
described by the Painlev\'e II equation when two clusters of eigenvalues 
get coalesced \cite{PII}, and more generally they 
are governed by the mKdV hierarchy \cite{mKdV}.

The recent use of matrix models for the study of gauge 
theory and string theory \cite{DV,DV2}
requires not only the knowledge of their 
critical behaviours but also their individual higher genus corrections 
away from criticality; the technology to compute them has been 
less developed in unitary 1-matrix models than in hermitian ones.
Moreover, some unconventional matrix models, which also have the Haar 
measure on the unitary group, have recently been proposed to describe 
a Chern-Simons theory \cite{Marino,AKMV} 
and five-dimensional gauge theories \cite{5d}. Therefore it is useful to develop computational techniques in these models as well. 

In this paper we show that, in fact, unitary and hermitian 1-matrix models 
are nothing but different descriptions of exactly the same 
system\footnote{In the double scaling limit, the duality of the two 
matrix models was established \cite{DJMW} more than a decade ago. 
The main point of the present paper is to show that the duality holds 
even {\it off criticality} so that one can use it to compute higher genus 
topological string amplitudes, etc.};  by a simple 
change of variables we map a general unitary 1-matrix model to a certain 
hermitian 1-matrix model coupled to $2N$ vectors with an appropriate 
potential. A similar transformation can also be done on the Chern-Simons 
matrix model. This means that all the techniques developed and results obtained 
in hermitian 1-matrix models ---  the higher genus computations of partition 
functions and correlation functions \cite{ACKM,Akemann,Chekhov,Eynard}, 
the relations to integrable hierarchies 
\cite{GMMMO,Martinec,AGL,BMS,Kostov}\footnote{
For unitary 1-matrix models it has been known that their underlying integrable 
structure is the ``modified Volterra'' \cite{GMMMO} or the 
``quarternionic Toda'' \cite{Martinec} hierarchy.} 
and conformal field theories \cite{CFT}, 
the derivation of the special geometry relation \cite{CSW,FM} --- 
they all can be used to investigate unitary 1-matrix models as well as other 
analogous models with the unitary measure.

The relevant transformation is simply the fractional linear transformation
\beqa
U&=&\frac{i1-\Phi}{i1+\Phi}, \label{duality}
\eeqa
where $U$ and  $\Phi$ are a unitary and a hermitian matrix, respectively.
The fact that their Haar measures are related by
\beqa
dU&=&\frac{d\Phi}{\det(1+\Phi^2)^N}
\eeqa
may be found in the book \cite{Hua} originally published in 1958. 
This transformation was used in \cite{BMS} to derive the Virasoro 
constraints \cite{Virasoro} 
in unitary 1-matrix models, although the 
fact that it implied an exact duality does not seem to have been fully recognized
\footnote{It was noted in the same paper (\cite{BMS}, Appendix D) that  
the question of whether the partition function of the full unitary model 
was a $\tau$ function of any integrable hierarchy remained open.} .
In \cite{GN} the possibility of using the transformation (\ref{duality}) 
was mentioned in the study of matrix models coupled to an external 
field, with a conclusion that it did not have ``a particularly attractive 
expansion'' for their purposes
\footnote{In \cite{GN} there was also mentioned another 
possibility $U=e^{i \Phi}$ \cite{Neuberger}, which was recently exploited 
by Tierz \cite{Tierz} in the study of the Chern-Simons matrix model. 
See section 5 for more details.\label{footnote:Tierz}}.
Later on, the techniques of computing 
the higher genus corrections were developed in hermitian 1-matrix 
models \cite{ACKM,Akemann,Chekhov,Eynard}, and more recently were
used in the Dijkgraaf-Vafa theory \cite{DV,DV2}. However, the transformation 
(\ref{duality}) has never been utilized in this context. 

The dual hermitian matrix models we end up with are closely related 
to what we encountered in the study of
four-dimensional ${\cal N}=1$ $U(n)$ gauge theories
coupled to $N_f$ fundamental matter fields (e.g. \cite{NSW,CSW}; see 
also \cite{FM} for further references); 
in the present case 
we will have vectors of order $N$ ($=2N$), 
however.
Nevertheless, we can take advantage of the derivation of the special 
geometry relation given in \cite{FM} to prove analogous relations
for unitary and the Chern-Simons matrix models.

The organization of this paper is as follows: In Section 2 we describe 
the details of the unitary/hermitian duality transformation, which plays 
a central role in the present paper, and demonstrate the use if it in 
the 1-cut Gross-Witten model as an illustration. In Section 3 we 
establish the relationship between the resolvent commonly used in 
unitary 1-matrix models and that in the dual hermitian models. We then 
derive the special geometry relation for general unitary 1-matrix models,
following \cite{FM}. Section 4 and 5 are devoted to the applications of 
the duality transformation. In Section 4 we use it in the unitary matrix 
formulation of the ${\cal N}=2$ pure $SU(2)$ gauge theory and prove 
that it correctly reproduces the genus-1 topological string amplitude,
whereas in Section 5 we deal with the matrix model mirror of 
Chern-Simons theory proposed in \cite{Marino}. Finally we conclude 
the paper with a summary. The two Appendices contain some technical 
details needed in the text.

\section{Unitary/Hermitian Duality}
\subsection{The Duality Transformation}
The partition function of the unitary 1-matrix model with a potential 
$W_U(U)$ is 
\beqa
Z&=&\frac1{{\rm Vol}U(N)} \int dU e^{-\frac1{g_s}{\rm Tr}W_U(U)}.
\eeqa
$U$ is an $N\times N$ unitary matrix and $dU$ is the Haar 
measure on $U(N)$. 
$g_s$ is the coupling constant. It is a classic result that this 
matrix integral is reduced to that over the eigenvalues \cite{HC} as
\beqa
Z&=&\int \prod_{j=1}^N d\theta_j e^{-\frac1{g_s}W_U(e^{i\theta_j})}
\prod_{k<l}^N\sin^2 \frac{\theta_k-\theta_l}2,
\label{Z}
\eeqa
where we take
$-\pi \leq \theta_j \leq \pi$  $(j=1,\ldots,N)$.

The partition function of the hermitian 
1-matrix model with a potential $W(\Phi)$ is given by 
\beqa
Z_{\mbox{\scriptsize hermitian}}
&=&\frac1{{\rm Vol}U(N)} \int d\Phi e^{-\frac1{g_s}{\rm Tr}W(\Phi)}\\
&=&\int \prod_{j=1}^N d\lambda_j e^{-\frac1{g_s}W(\lambda_j)}
\prod_{k<l}^N (\lambda_k-\lambda_l)^2.
\label{Zhermitian}
\eeqa
$\Phi$ is an $N\times N$ hermitian matrix and 
$\lambda_j$ $(j=1,\ldots,N)$
are the eigenvalues of $\Phi$.

In hermitian models the Jacobian associated with the gauge 
fixing is known to be a Vandermonde determinant; 
the $\log$ of it may be viewed as a 
potential for a repulsive force between two eigenvalues on the real axis. For unitary models, on the other hand, $\sin$ replaces the linear function in the Jacobian, 
so that it may also be viewed as representing the same linear force between 
two eigenvalues of $U$, located this time on a unit circle because
\beqa
\sin^2 \frac{\theta_k-\theta_l}2
&=&\frac14 |e^{i\theta_k}-e^{i\theta_l}|^2.
\eeqa

To write $Z$ as $Z_{\mbox{\scriptsize hermitian}}$ for some $W(\Phi)$ 
we have only to use the fractional linear transformation 
\beqa
w&=&\frac{i-z}{i+z}
\label{w}
\eeqa
on the complex plane,
which maps the unit circle $|w|=1$ to the real axis $z\in{\bf R}$. If
$w=e^{i\theta}$, it is equivalent to 
\beqa
z=\tan\frac\theta 2.
\eeqa
With this elementary change of variables the determinant factor  
becomes 
\beqa
\prod_{k<l}^N\sin^2 \frac{\theta_k-\theta_l}2
&=&
\prod_{j=1}^N \frac1{(1+\lambda_j^2)^{N-1}}
\prod_{k<l}^N (\lambda_k-\lambda_l)^2.
\eeqa
Since 
\beqa
d\theta_j&=&\frac{2d\lambda_j}{1+\lambda_j^2}
\eeqa
has another factor of $\frac1{1+\lambda_j^2}$,
we find 
\beqa
Z&=&2^N
\int \prod_{j=1}^N d\lambda_j 
e^{-\frac1{g_s}W_U\left(\frac{i-\lambda_j}{i+\lambda_j}\right) 
-N\log(1+\lambda_j^2)}
\prod_{k<l}^N (\lambda_k-\lambda_l)^2.
\label{Zafter}
\eeqa
Therefore, in the 't Hooft expansion \cite{tHooft} with 
\beqa
\mu&\equiv&g_s N
\eeqa
kept fixed, $N$ factors out in the exponential, leaving a hermitian matrix model 
with the potential 
\beqa
W(\lambda)&=&W_U\left(\frac{i-\lambda}{i+\lambda}\right) +\mu \log(1+\lambda^2).
\label{W}
\eeqa

$Z$ (\ref{Zafter}) may be thought of as arising from a hermitian 
matrix model coupled to vectors:
\beqa
Z_{\mbox{\scriptsize +vector}}
&=&\frac1{{\rm Vol}U(N)} \int d\Phi 
\prod_{I=1}^{N_f} dQ_I dQ^{\dagger}_I
\exp\left[-\frac1{g_s}\left({\rm Tr}W_\Phi(\Phi)
+\sum_{I=1}^{N_f} (Q_I^\dagger \Phi Q^I 
-m_I Q_I^\dagger Q^I)\right)\right]
\nonumber\\
&\propto& \int d\Phi 
e^{-\frac1{g_s}\left({\rm Tr}W_\Phi(\Phi)
+g_s \sum_{I=1}^{N_f} \log(\Phi-m_I)\right)}
\label{Z+vector}
\eeqa
after integrating out $Q_I$, $Q^{\dagger}_I$.
Since
\beqa
\mu \log(1+\lambda^2)&=&g_s N\left(\rule{0mm}{5mm}
\log(\lambda+i) +\log(\lambda-i)
\right),
\eeqa
we see that the unitary 1-matrix model partition function $Z$ (\ref{Z}) is, 
up to an $N$-independent multiplicative constant, 
equal to that of the hermitian matrix model $Z_{\mbox{\scriptsize +vector}}$ 
(\ref{Z+vector}) with a potential 
\beqa
W_\Phi(\Phi)&=&W_U\left(\frac{i1-\Phi}{i1+\Phi}\right), 
\eeqa
coupled to $N_f=2N$ vectors having pure imaginary ``masses'' $\pm i$.

Some remarks are in order here: 
\begin{itemize}
\item[1.]{The potential $W(\lambda)$ (\ref{W}) does not depend on $N$.
Therefore the equivalence exactly holds to all orders in the $1/N$ expansion.}
\item[2.]{The extra $\log$ term in $W(\lambda)$ can be expanded 
as a Taylor series near $\lambda=0$. Thus all the techniques developed 
for the computations of higher genus amplitudes \cite{ACKM,Akemann,Chekhov} can also be applied 
\footnote{as far as $W_U(U)$ is regular at $U=1$. For example, 
$W_U(U)$ can be any polynomial of $U$ and $U^{-1}$.}
to those of unitary 1-matrix models. This is in contrast to the transformation
of the type such as $U=\exp i\Phi$, 
which leads to a logarithmic singularity at $\lambda=0$.}
\item[3.]{An immediate consequence of the duality is that the partition function 
of a unitary 1-matrix model is also a $\tau$-function of the KP hierarchy.}
\item[4.]$Z_{+\mbox{\scriptsize vector}}$ (\ref{Z+vector})
is the very matrix model used for computing the effective superpotential 
and other $F$-terms of a four-dimensional ${\cal N}=1$ $U(n)$ gauge theory
coupled to $N_f$ fundamental matter fields (e.g.\cite{NSW,CSW}), although 
$N_f=2N>2n$ is the region where this interpretation breaks down.
\end{itemize}

\subsection{Example: The Gross-Witten Model}
To demonstrate the use of the duality transformation we will first 
rederive the well-known 1-cut solution of the Gross-Witten model \cite{GW}
by using standard hermitian matrix model techniques \cite{BIPZ}. 
The Gross-Witten model is characterized by the potential
\beqa
W_U(U)&=&-\left(
U+\frac1U
\right)
~=~-2\cos\theta~~~~~~(U=e^{i\theta}).
\label{WGrossWitten}
\eeqa
The corresponding potential of the hermitian matrix model is 
\beqa
W(z)&=&-\frac{2(1-z^2)}{1+z^2}+\mu\log(1+z^2).
\eeqa
The resolvent is given by
\beqa
\omega(z)&=&\frac{\sqrt{(z-a)(z-b)}}{4\pi i \mu}
\oint_A dx \frac{W'(x)}{(z-x)\sqrt{(x-a)(x-b)}},
\label{omega1cut}
\eeqa
where $a$,$b$ are the end points of the cut satisfying
\beqa
\oint_A dx \frac{W'(x)}{\sqrt{(x-a)(x-b)}}&=&0,\nonumber\\
\oint_A dx \frac{xW'(x)}{\sqrt{(x-a)(x-b)}}&=&4\pi i \mu.
\label{1cutconditions}
\eeqa
The contour $A$ surrounds the cut, running close enough to it 
so that the contour avoids the branch cuts of $\log(1+z^2)$. 
The conditions (\ref{1cutconditions}) follow from the large $|z|$ 
behavior of $\omega(z)$. They are easily solved to give
\beqa
b=-a=\sqrt{\frac\mu{2-\mu}}~(>0),
\eeqa
so that
\beqa
\omega(z)&=&\frac1{2\mu}\left(
W'(z)-\frac{8\sqrt{z^2-b^2}}{\sqrt{1+b^2}(1+z^2)^2}
\right).
\eeqa
The spectral density $\rho(\lambda)$ may be read off from the 
discontinuity of the resolvent $\omega(z)$ as
\beqa
\rho(\lambda)d\lambda
&=&-\frac1{2\pi i}(\omega(\lambda+i0)-\omega(\lambda-i0))
d\lambda
\nonumber\\
&=&
{\displaystyle \frac4{\pi \mu}
\frac{\sqrt{b^2-\lambda^2}}{\sqrt{1+b^2}(1+\lambda^2)^2}
d\lambda
}\nonumber\\
&=&{\displaystyle \frac2{\pi \mu}\cos\frac\theta 2
\sqrt{\frac\mu 2-\sin^2\frac\theta 2}d\theta},
\eeqa
which is precisely the solution of the Gross-Witten model 
in the weak-coupling regime \cite{GW}.

The genus-0 free energy $F_0$ can be obtained as that of the 
hermitian matrix model \cite{BIPZ}:
\beqa
F_0&=&-\frac\mu 2\int_{-\infty}^\infty
d\lambda \rho(\lambda)W(\lambda)
+\mu^2 \int_{-\infty}^\infty d\lambda \rho(\lambda)
\log|\lambda|
-\frac
\mu2 W(0).
\eeqa
The genus-1 free energy $F_1$ is also known to be given by 
\cite{ACKM}
\beqa
F_1&=& -\frac1{24}\left(
\log M_1 + \log J_1 +4 \log 2b
\right)
\eeqa
with
\beqa
M_1&=&\oint_A \frac{dw}{2 \pi i}\frac{\frac1\mu W'(w)}
{\sqrt{w^2-b^2}(w+b)},\nonumber\\
J_1&=&\oint_A \frac{dw}{2 \pi i}\frac{\frac1\mu W'(w)}
{\sqrt{w^2-b^2}(w-b)}.
\eeqa
Straightforward residue computations yield
\beqa
M_1&=&\displaystyle\oint_A \frac{dw}{2 \pi i}\frac{
\frac{8w}{\mu(w+i)^2(w-i)^2 }+
\frac{2w}{(w+i)(w-i)}
}
{\sqrt{w^2-b^2}(w+b)}\\
&=&\frac4{b^2\sqrt{1+b^2}^3}~~~=~J_1.
\eeqa
Therefore
\beqa
F_1&=&\displaystyle\frac18\log(1+b^2)-\frac13\log 2\nonumber\\
&=&\frac18\log(2-\mu)-\frac5{24}\log 2,
\eeqa
which also agrees with the known result of \cite{GN} up to a constant.

\section{Resolvents and The Special Geometry Relation}
\subsection{Resolvents in Unitary/Hermitian Matrix Models}
In this subsection we will examine the difference
between the two resolvents:  one is that commonly used in 
unitary 1-matrix models, and the other is the usual resolvent in 
hermitian models.
The saddle point equation derived from (\ref{Z}) is
\beqa
\frac1N\sum_{k\neq j}^N\cot\frac{
\theta_j-\theta_k}2~=~
\frac i\mu e^{i\theta_j}W'_U(e^{i\theta_j}).
\label{unitarysaddlept}
\eeqa
The resolvent $\omega_U(x)$ is customarily defined in unitary 
matrix models as
\beqa
\omega_U(x)&\equiv&\frac1N \sum_{j=1}^N
\cot \frac{x-\theta_j}2~~
\stackrel{N\rightarrow \infty}{\rightarrow}
\int_{-\pi}^\pi d\theta \rho_U(\theta)
\cot\frac{x-\theta}2.
\eeqa
On the other hand, the resolvent in hermitian 1-matrix  
models is given by
\beqa
\omega(z)&\equiv&\frac1N \sum_{j=1}^N
\frac1{z-\lambda_j}~~
\stackrel{N\rightarrow \infty}{\rightarrow}
\int_{-\infty}^\infty d\lambda \rho(\lambda)
\frac1{z-\lambda}.
\eeqa
The relation between the two is as follows: The spectral densities 
satisfy
\beqa
d\theta \rho_U(\theta)&=&d\lambda \rho(\lambda).
\eeqa
It is easy to show that as a differential 
\beqa
dx \cot\frac{x-\theta}2&=&
dz\left(
\frac2{z-\lambda}-\frac{2z}{z^2+1}
\right),
\eeqa
where 
$z=\tan\frac x2$ and
$\lambda=\tan\frac\theta 2$.
Thus we obtain the relation
\beqa
\omega_U(x)dx&=&\left(2\omega(z)-(\mu\log(1+z^2))'\right)dz.
\eeqa
If we define
\beqa
y_U(x)&\equiv& \frac d{dx}W_U(e^{ix})-\mu\omega_U(x),\nonumber\\
y(z)&\equiv& W'(z)-2\mu\omega(z),
\eeqa
then we find
\beqa
y_U(x)dx&=&y(z)dz.
\eeqa
Therefore, the function $y_U(x)$ (or rather the differential $y_U(x)dx$), 
which characterizes the curve of the unitary matrix model, is 
simply computed as $y(z)$ (or $y(z)dz$) in the corresponding 
hermitian matrix model. 

\subsection{The Special Geometry Relation}
One of the properties of matrix models that makes contact with gauge 
theory and topological string theory is the ``special geometry'' relation \cite{DV},
a relation between the genus-0 free energy $F_0$ and 
periods of the differential $y(z)dz$. It is due to this property that makes 
possible to identify $F_0$ as a prepotential of special geometry \cite{special}. 

This relation was recognized in \cite{DV}.
For hermitian 1-matrix models with a polynomial potential, a proof using 
a Legendre transformation was given in \cite{CSW} and a more direct one 
in \cite{FM}.  Also for unitary 1-matrix models, one may argue that 
a similar relation holds \cite{DV2} because $y(x)$ is again interpreted 
as a ``force'' on an eigenvalue exerted by other ones located on the unit 
circle. Strictly speaking however, the relevant period integrals diverge 
without a regularization, the detail of which is not determined by the 
intuitive argument. In this subsection we will derive, following (and 
appropriately modifying) the proof in \cite{FM}, the special geometry 
relation for unitary 1-matrix models.

Let $W_U(U)$ be any polynomial of $U$ and $U^{-1}$.
The genus-0 free energy for $n$-cut solutions is given by 
\beqa
F_0&=&-\frac\mu 2\int_{-\infty}^\infty
d\lambda \rho(\lambda)W(\lambda)
+\mu\sum_{j=1}^n \mu_k\int_{-\infty}^\infty d\lambda \rho(\lambda)
\log|\lambda-\lambda_{0j}|
-\frac12\sum_{j=1}^n
\mu_j W(\lambda_{0j}),
\label{F0multicut}
\eeqa
with $W(\lambda)$ given by (\ref{W}), and 
\beqa
\mu_j&\equiv&\mu\int_{a_{2j-1}}^{a_{2j}}d\lambda \rho(\lambda)
~~~(j=1,\ldots,n).
\label{muj}
\eeqa
$a_{2j-1}$, $a_{2j}$ are the two end points of the $j$th cut, whereas
$\lambda_{0j}$ is an arbitrary point on the $j$th cut, for $j=1,\ldots,n$. 
We assume that they are all real and $a_1<a_2<\cdots<a_{2n-1}<a_{2n}$.  
This expression for the free energy is more or less standard, but we 
give a derivation in Appendix A for completeness. 

The right hand side of (\ref{F0multicut})
can be written in terms of contour integrals as
\beqa
F_0&=&\frac1{8\pi i}\sum_{j=1}^n \oint_{A_j}dz y(z) W(z) 
+\lim_{\Lambda_0\rightarrow -\infty}
\left[
\sum_{j=1}^n \frac{\mu_j}4 \int_{\hat B_j} dz y(z) ~-\frac\mu 2 W(\Lambda_0)
+\mu^2 \log|\Lambda_0|
\right].\nonumber \\
\eeqa
The definitions of the contours are shown in Figure 1. 

\begin{figure}
\centering
\resizebox{0.65\textwidth}{!}{
  \rotatebox{0}{\includegraphics{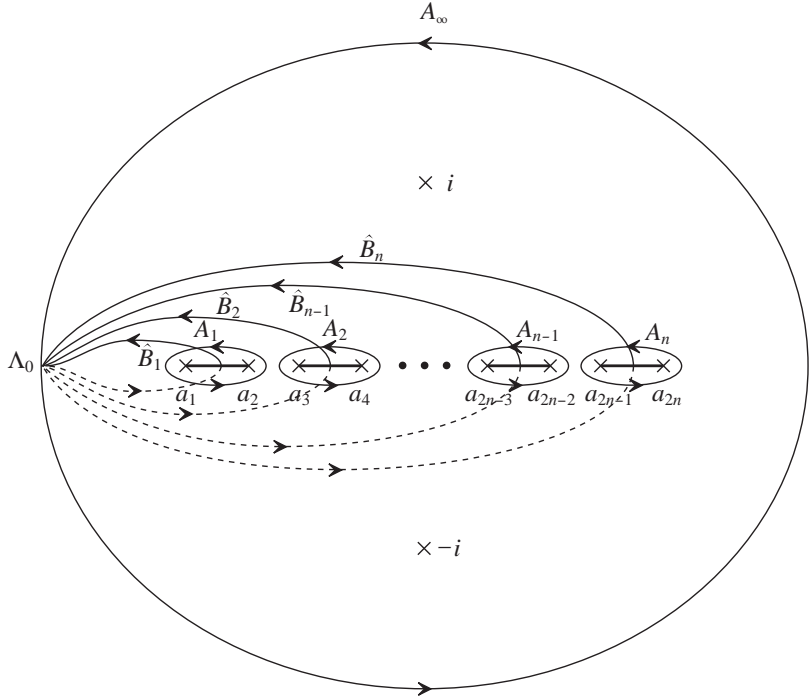}}}
\caption{The contours. Those 
on the second sheet are shown in broken lines.
 }
\label{contour1}      
\end{figure}

$y(z)$ is given by the equation 
\beqa
y(z)
&=&W'(z)-2\mu\omega(z)\nonumber\\
&=& -\frac{\sqrt{\prod_{k=1}^{2n}
(z-a_k)}}{2\pi i }
\left(\oint_{A_\infty}-\oint_{x=i}-\oint_{x=-i}\right) dx 
\frac{W'(x)}{(z-x)\sqrt{\prod_{k=1}^{2n}(x-a_k)}},
\label{y(z)}
\eeqa
so that
\beqa
\oint_{A_j} dz y(z)~=~-4 \pi i \mu_j ~~(j=1,\ldots,n).
\label{yAperiod}
\eeqa

Our aim is to express $\frac{\partial F_0}{\partial \mu_j}$ 
in terms of the periods of $y(z)dz$.
For this purpose we need to compute $\frac{\partial y(z)}{\partial \mu_j}dz$,
which may be found by simply studying the locations of the singularities,
as was done in \cite{FM}. In the present case, however, $W'(z)$ has 
extra pole singularities at $z=\pm i$ with residue $\mu$ coming from 
the unitary Haar measure and must be also taken into account.

$y(z)dz$ has the following properties:
\begin{itemize}
\item{Since the contours of (\ref{y(z)}) do not enclose $x=z$, 
the integrals give single-valued functions of $z$ on the complex plane. 
Therefore $y(z)$ reverses its sign under the exchange of the first and 
second sheets.}
\item{Since we have assumed that $W_U(U)$ is a polynomial of 
$U$ and $U^{-1}$, $y(z)dz$ is regular at $z=\infty,\tilde\infty$, where we 
denote here and henceforth a point on the second sheet by its value of 
$z$ with tilde.}
\item{Since $\omega(z)dz$ must be regular at $z=\pm i$ on the first sheet, 
$y(z)dz$ exhibits the same singular behavior as $W'(z)dz$ at $z=\pm i$, 
and as $-W'(z)dz$ at $z=\widetilde{\pm i}$.}
\end{itemize}
These properties, together with the condition (\ref{yAperiod}), uniquely 
determine the $\mu_j$ derivative of $y(z)dz$ as 
\beqa
\frac{\partial y(z)}{\partial \mu_j}dz~=~-4\pi i~ \hat\omega_j(z),
\eeqa
where
\beqa
\hat\omega_j&=&\omega_j+\hat\omega_n~~~
(j=1,\ldots,n-1),\nonumber\\
\hat\omega_n&=&-\frac1{4 \pi i}
\left(\omega_{(+i)-\widetilde{(+i)}}+
\omega_{(-i)-\widetilde{(-i)}}\right).
\eeqa
$\omega_j$'s are holomorphic differentials normalized as
\beqa
\oint_{A_i}\omega_j&=&\delta_{ij}~~~(j=1,\ldots,n-1),
\eeqa
and $\omega_{a-b}(z)$ denotes a zero 
A-period abelian differential of the third kind having simple poles 
at $z=a,b$ with residue $\pm1$, respectively. 

Having computed $\frac{\partial y(z)}{\partial \mu_j}dz$, we are now able 
to derive the special geometry relation. Differentiating the free 
energy $F_0$ by $\mu_j$, we have.
\beqa
\frac{\partial F_0}{\partial \mu_i}
&=&-\frac12\sum_{j=1}^n \oint_{A_j}\hat\omega_i(z) W(z) 
+\frac1{8\pi i}\sum_{j=1}^n \oint_{A_j}dz y(z) 
\log(z^2+1)
\nonumber
\\
&&+\lim_{\Lambda_0\rightarrow -\infty}
\left[-\pi i 
\sum_{j=1}^n \mu_j \int_{\hat B_j} \omega_i(z) 
~+\frac14 \int_{\hat B_i} dz y(z)
~-\frac12 W(\Lambda_0) \right.\nonumber\\ && \left.
-\frac\mu 2\log(\Lambda_0^2+1)
~+2\mu \log|\Lambda_0|
\right]
\nonumber\\
&=&-\frac12\sum_{j=1}^n \oint_{A_j}\hat\omega_i(z) W(z) 
~+\lim_{\Lambda_0\rightarrow -\infty}
\left[-\pi i 
\sum_{j=1}^n \mu_j \int_{\hat B_j} \omega_i(z) 
\right.\nonumber\\ 
&&\left.
+\lim_{\epsilon\rightarrow 0}
\left(
\frac14 \left(
\int_{\hat B_i} \!\!\!+\int_{\Lambda_0}^{i-\epsilon}
\!\!\!+\int_{\Lambda_0}^{-i-\epsilon}
\right)dz y(z)
~-W(i-\epsilon)-W(-i-\epsilon)\right)
\rule{0mm}{7mm}
\right].
\eeqa
This expression can be simplified because 
one can show that the following relation holds:
\beqa
&&-\frac12\sum_{j=1}^n \oint_{A_j}\hat\omega_i(z) W(z) 
~+\lim_{\Lambda_0\rightarrow -\infty}
\left[-\pi i 
\sum_{j=1}^n \mu_j \int_{\hat B_j} \omega_i(z) 
\right]\nonumber\\ 
&=&
\lim_{\Lambda_0\rightarrow -\infty}
\left[\frac14 \lim_{\epsilon\rightarrow 0}
\left(
\int_{\hat B_i} \!\!\!+\int_{\Lambda_0}^{i-\epsilon}
\!\!\!+\int_{\Lambda_0}^{-i-\epsilon}
\right)dz y(z)
~-W(i-\epsilon)-W(-i-\epsilon)
\right].
\label{Riemann'srelation}
\eeqa
This relation may be derived in the way parallel to that in \cite{FM} 
by using Riemann's bilinear relation; we give a brief sketch of it 
in Appendix B. Thus we obtain the final result 
\beqa
\displaystyle\hskip -48pt \frac{\partial F_0}{\partial \mu_i}
&=&
\frac12 \lim_{\epsilon\rightarrow 0}\left[
 \lim_{\Lambda_0\rightarrow -\infty}
\left(
\int_{\hat B_i} \!\!\!+\int_{\Lambda_0}^{i-\epsilon}
\!\!\!+\int_{\Lambda_0}^{-i-\epsilon}
\right)dz y(z)
~-W(i-\epsilon)-W(-i-\epsilon)\right]\nonumber\\
&=&
\frac12 \lim_{\epsilon\rightarrow 0}\left[
\int_{\widetilde{-i+i\epsilon}} ^{i-i\epsilon}
\!\!\!\!\!\!\!\!\!\mbox{\scriptsize ($i$th cut) }
dz y(z)
~-W(i-i\epsilon)-W(\widetilde{-i+i\epsilon})\right],
\eeqa
where in the last line the integration contour passes through the 
$i$th cut. If we go back to the $w$-plane (\ref{w}), 
the contour runs from a point infinitely close to $w=0$ to another 
to $w=\infty$.
\section{Application I : $N=2$ Gauge Theory via Unitary Matrix Models}
It is now well-known that the effective superpotential and other F-terms in 
${\cal N}=1$ supersymmetric gauge theory can be calculated using 
hermitian matrix models \cite{DV,DV2}. The ${\cal N}=2$ limit may also 
be achieved by turning off the superpotential with  
the effective superpotential kept extremized with respect to the glueball 
fields \cite{CV}. On the other hand, it has been shown \cite{DV2} that 
there is another route to realize more directly the pure 
$SU(2)$ Seiberg-Witten curve via a unitary matrix model 
without referring to any ${\cal N}=1$ theory. 
In this section we will apply the duality transformation to this unitary 
matrix model to examine whether the genus-1
topological amplitude, obtained by other means, is correctly reproduced 
in this framework.
\subsection{2-cut Solutions of The Gross-Witten Model}
To realize the pure $SU(2)$ Seiberg-Witten curve we again consider 
the Gross-Witten model, though this time focus on 2-cut solutions. 
In this case it is convenient to shift $\theta$ by 90 degrees 
in (\ref{WGrossWitten}) and consider the potential 
\beqa
W_U(e^{i\theta})&=&-2\sin\theta,\\
W(z)&=&\frac{4z}{1+z^2}+\mu\log(1+z^2).
\eeqa
We let $a_1<\cdots<a_4$ be the end points of the cuts; they are 
constrained by 
\beqa
\left(\oint_{A_1}+\oint_{A_2}\right) dx \frac{W'(x)}{\sqrt{\prod_{k=1}^{4}(x-a_k)}}
&=&0,
\nonumber\\
\left(\oint_{A_1}+\oint_{A_2}\right) dx \frac{xW'(x)}{\sqrt{\prod_{k=1}^{4}(x-a_k)}}
&=&0,
\label{2cutendpoints}
\\
\left(\oint_{A_1}+\oint_{A_2}\right) dx \frac{x^2W'(x)}{\sqrt{\prod_{k=1}^{4}(x-a_k)}}
&=&4\pi i \mu.
\nonumber
\eeqa
One can straightforwardly compute the differential $y(z)dz$ to find
\beqa
y(z)&=& -\frac{\sqrt{\prod_{k=1}^{4}
(z-a_k)}}{2\pi i }
\left(\oint_{A_\infty}-\oint_{x=i}-\oint_{x=-i}\right) dx \frac{W'(x)}{(z-x)\sqrt{\prod_{k=1}^{4}(x-a_k)}}\nonumber\\
&=&\frac8c \frac{\sqrt{\prod_{k=1}^{4}
(z-a_k)}}{(z^2+1)^2},\label{y2cut}
\eeqa
where we put
$\sqrt{\prod_{k=1}^{4}(i-a_k)}=
\sqrt{\prod_{k=1}^{4}(-i-a_k)}\equiv c$.

To incorporate the end-point conditions (\ref{2cutendpoints}) 
we go back to the $w$-plane (\ref{w}). The two cuts are mapped 
to two arcs located on the unit circle with end points given by  
\beqa
T_k=\frac{i-a_k}{i+a_k}~~~(k=1,\ldots,4).
\eeqa
Then the conditions (\ref{2cutendpoints}) are fulfilled if and only if 
$T_k$'s are the roots of the following equation of $U$:
\beqa
U^4-2i\mu U^3+\sigma_2 U^2 +2i\mu U +1=0,
\eeqa
where
\beqa
\sigma_2\equiv T_1T_2+T_1T_3+T_1T_4+T_2T_3+T_2T_4+T_3T_4
\eeqa
is a free parameter, which is eventually identified as (essentially) 
the $u$ modulus of the Seiberg-Witten curve. Using the relations between the roots 
and the coefficients, $y(z)$ (\ref{y2cut}) may be written in the form 
obtained in \cite{DV2}:
\beqa
y(z) dz=2\sqrt{
\cos^2\theta +\mu(\sin \theta -u)
}~d\theta
\eeqa
with
$u\equiv \frac{2-\sigma_2}{4\mu}$.
The ${\cal N}=2$ curve and differential are obtained by restricting the 
moduli to a co-dimension one sublocus in the following way: We first scale 
$y(z)dz$ as
\beqa
y(z)dz&\equiv&2\sqrt{\mu}~ y_{\mbox{\scriptsize SW}}(\theta)d\theta
\eeqa
and take the $\mu\rightarrow\infty$ limit. Then we obtain the Seiberg-Witten 
curve for the pure $SU(2)$ gauge theory:
\beqa
y_{\mbox{\scriptsize SW}}^2-\sin \theta+u&=&0.
\label{ySW}
\eeqa
On the other hand, we 
choose, as was done in \cite{CV}, $\mu_1$ and $\mu(=\mu_1+\mu_2)$ 
(instead of $\mu_1$ and $\mu_2$) as 
independent parameters. 
Using the special geometry relation in the previous section, 
we have
\beqa
\left.\frac{\partial F_0}{\partial \mu_1}
\right|_{\mbox{\scriptsize $\mu$}}&=&
\left.\frac{\partial F_0}{\partial \mu_1}
\right|_{\mbox{\scriptsize $\mu_2$}}-
\left.\frac{\partial F_0}{\partial \mu_2}
\right|_{\mbox{\scriptsize $\mu_1$}}\nonumber\\
&=& -\frac12 \oint_B dz y(z),
\eeqa
where $B=\hat B_2 \hat B_1^{-1}$.
Therefore $y_{\mbox{\scriptsize SW}}(\theta)d\theta$ is indeed a 
Seiberg-Witten differential.

\subsection{Gravitational $F$-terms via The Unitary Matrix Model}
We are now interested in the genus-1 free energy of this 
matrix model. In the 2-cut case it is given by \cite{Akemann}
\beqa
F_1=-\frac1{24}\log\prod_{j=1}^4 M_j
-\frac12\log |K(k)|
-\frac1{12}\log\prod_{i<j}(a_i-a_j)^2
+\frac14\log(|a_1-a_3||a_2-a_4|),
\label{F1twocut}
\eeqa
where the moments are defined by
\beqa
M_i&=&\frac1{2\pi i}\left(\oint_{A_1}\!\!\!+\oint_{A_2}\right) dx 
\frac{\frac1\mu W'(x)}{(x-a_i)\sqrt{\prod_{k=1}^{4}(x-a_k)}},
\eeqa
and
\beqa
K(k)=\int_0^1 dt \frac1{\sqrt{(1-t^2)(1-k^2t^2)}}
\eeqa
is the complete elliptic integral of the first kind
with modulus
\beqa
k^2~=~\frac{(a_1-a_4)(a_2-a_3)}{(a_1-a_3)(a_2-a_4)}
~=~\frac{2\sqrt{u^2-1}}{u+\sqrt{u^2-1}}.
\eeqa
Here we note that $y_{SW}^2$ (\ref{ySW}) has at most only two 
zeros on the unit circle, and hence is only defined as an 
analytic continuation of the 2-cut solutions. Therefore we evaluate 
$M_i$'s where it is well-defined, and then take the 
$\mu\rightarrow\infty$ limit afterwards. This is very different from the 
situation in the hermitian matrix realization \cite{KMTDST} where 
a spectral curve ``really'' exists in the ${\cal N}=2$ limit. 
Nevertheless, despite the difference, $F_1$ (\ref{F1twocut}) obtained 
as an analytic continuation correctly reproduces the known result, 
as we will see below.

$M_1$ is found after some straightforward residue computations as
\beqa
M_1&=&-\frac2{\mu c^3(a_1^2+1)}
\left[\rule{0mm}{4mm}
-8 +6(a_2a_3 + a_2a_4 + a_3a_4)
-3a_1(a_2 + a_3 + a_4 - a_2a_3a_4)\right.
\nonumber\\ &&\left.
\hskip 15mm
+a_1^2(-5 +3(a_2a_3 + a_2a_4 + a_3a_4))
\right]
-\frac{2(-\mu a_1 +2)}{\mu c(a_1^2+1)}
\eeqa
with 
$c^2=\frac{16}{\sigma_2+2}$. Various symmetric polynomials of 
$a_i$'s turn out to be
\beqa
a_1 +a_2 +a_3 +a_4&=&\frac{2\mu}{\mu u -1},\nonumber \\
a_1a_2 + a_1a_3 + a_1a_4+a_2a_3 + a_2a_4 + a_3a_4
&=&2+\frac4{\mu u-1},\nonumber \\
a_1a_2a_3 +a_1a_2a_4+a_1a_3a_4+a_2a_3a_4
&=&\frac{2\mu}{\mu u -1},\nonumber \\
a_1a_2a_3a_4 &=&1.\rule{0mm}{5mm}
\eeqa
Thus we find
\beqa
\prod_{j=1}^4 M_j=\frac{(1-\mu u)^6}{\mu^4}
~=~\mu^2 u^6\left(1+{\cal O}(\mu^{-1})\right).
\eeqa
The discriminant can be computed as 
\beqa
\prod_{i<j}(a_i-a_j)^2
&=&\frac{16^2 \mu^2 (\mu^2-4\mu u+4)^2 (u^2 -1)}
{(\mu u-1)^6}
\nonumber\\
&\stackrel{\mu\rightarrow\infty}
{\longrightarrow}&
\frac{16^2(u^2-1)}{u^6},
\eeqa
and also
\beqa
|a_1-a_3||a_2-a_4|&=&
2\left(
1+\sqrt{1-\frac1{u^2}}
\right).
\eeqa
Plugging these data into (\ref{F1twocut}), we see that 
the $\log u$ terms precisely cancel. 
This leaves
\beqa
F_1
&=&
-\frac12\log K(k)
-\frac1{12}\log(u^2-1)
+\frac14\log(u+\sqrt{u^2-1})
-\frac1{12}\log2\mu 
+{\cal O}(\mu^{-1}),
\nonumber\\
\eeqa
the $u$-dependent part of which is in agreement with the known 
genus-1 result \cite{MW,KMTDST}. 

\section{Application II : A Matrix Model Mirror of A Chern-Simons Theory}
\subsection{The Transformation}
The final example we consider is a matrix model mirror of a Chern-Simons 
theory. It has been proposed \cite{Marino} that the partition function of the 
$SU(N)$ Chern-Simons theory of $S^3$ is equal to the partition function 
of a matrix model with a ``Wick-rotated'' Jacobian as 
\beqa
Z_{\mbox{\scriptsize CS}}&=&e^{-\frac\mu{12}(N^2-1)} Z,\nonumber\\
Z
&=&\int \prod_{j=1}^N du_j e^{-\frac1{g_s}
\frac{u_j^2}2}
\prod_{k<l}^N\sinh^2 \frac{u_k-u_l}2.
\label{ZCS}
\eeqa
In this section we apply the duality transformation to this model, compute 
the genus-0 and~-1 free energies and confirm that they indeed reproduce
those of the Chern-Simons theory. In fact, it was already proven by Tierz 
\cite{Tierz} that the partition function (\ref{ZCS}) is equal 
to the Chern-Simons partition function to all orders in the $1/N$ expansion,
by utilizing yet another transformation (as in footnote \ref{footnote:Tierz})
to map (\ref{ZCS}) to that of a 
hermitian matrix model (different from ours) and use its orthogonal 
polynomials. However, there are still some benefits in considering 
{\it our} transformation in this model as follows:  
First, it will also demonstrate how to compute explicitly the lower genus 
free energies in other such models that have the same measure of the 
``Wick-rotated'' type, such as those proposed to be related to 
some five-dimensional gauge theories \cite{5d}, even where no good 
orthogonal polynomials are available. Secondly, because our 
transformation induces no singularity near the origin in the potential, 
one can show the special geometry relation in these models by only
slightly modifying the proof for hermitian models. Finally, for the 
same reason, the relation to the KP hierarchy is made manifest by our 
transformation, and not by the one adopted in \cite{Tierz} because of its
$\log$ branch at the origin. 

For the measure of the type (\ref{ZCS}) the transformation we need is
\beqa
\lambda_j&=&\tanh\frac{u_j}2.
\label{CStransformation}
\eeqa
$Z$ (\ref{ZCS}) can then be written as
\beqa
Z
~=~2^N\int \prod_{j=1}^N d\lambda_j 
e^{-\frac N{\mu}\left[\frac12\left( \log\frac{1+\lambda_j}{1-\lambda_j}\right)^2
+\mu\log(1-\lambda_j^2)
\right]}
\prod_{k<l}^N (\lambda_k-\lambda_l)^2. 
\eeqa
Comparing it with (\ref{Z+vector}),
we find that the Chern-Simons matrix model has,
up to an $N$-independent multiplicative constant, 
the same partition function as 
that of the hermitian matrix model with a potential 
\beqa
W(z)= \frac12\left(\log\frac{1+z}{1-z}\right)^2+\mu\log(1-z^2),
\label{WCS}
\eeqa
or as that of the hermitian matrix model 
coupled to $N_f=2N$ vectors having {\it real} masses $\pm 1$
with a potential 
\beqa
W_\Phi(z)= \frac12\left(\log\frac{1+z}{1-z}\right)^2.
\eeqa
In this case, therefore, the potential of the resulting dual hermitian 
matrix model has branch cuts of $\log$ ending on the real axis; 
this is not a problem because the eigenvalues are located only 
on the interval between 1 and $-1$, as is clear from the transformation 
(\ref{CStransformation}).

\subsection{The Spectral Curve}
We will now rederive the (nontrivial part of the) geometry of
the deformed conifold as the spectral curve, following the standard 
procedure as before. Let us consider a 1-cut solution with a cut $[a,b]$. 
The end point conditions (\ref{1cutconditions}) with $W$ (\ref{WCS}) 
imply 
\beqa
b=-a=\sqrt{1-e^{-\mu}}~~~(>0).
\eeqa
The resolvent is also given by (\ref{omega1cut}) with $W$ (\ref{WCS}). 
The contribution of the second term of $W$ is simply obtained by residue 
computations, while the contribution of the first term can be written as
\beqa
\frac{(z^2-b^2)^{\frac12}}{4\pi i\mu}(I(z)+I(-z)),
\eeqa 
where
\beqa
I(z)&\equiv&\oint_A dx \frac{2\log (1+x)}
{(z-x)(1-x^2)(x^2-b^2)^{\frac12}}.
\eeqa
If the contour $A$ is deformed to a large circle around infinity, $I(z)$ 
also picks up the contribution from the difference of the integrals along 
both sides of the branch cut of $\log (1+x)$. After some calculations 
we find 
\beqa
I(z)&=&\frac{4\pi i}{(1-z^2)\sqrt{z^2-b^2}}
\log\frac{z+b^2+\sqrt{(z^2-b^2)(1-b^2)}}{z+\sqrt{z^2-b^2}}
\nonumber\\
&&+\frac{2\pi i}{\sqrt{1-b^2}}\left(
\frac{\log(1+\sqrt{1-b^2})}{1-z}
-\frac{1}{1+z} \log\frac{1+\sqrt{1-b^2}}{2(1-b^2)}
-\frac{\log 2}{1-z} 
\right),
\eeqa
and
\beqa
\omega(z)&=&\frac1{\mu(1-z^2)}
\left[
\log\frac{z^2+b^2+\sqrt{(z^2-b^2)(1-b^2)}}{z^2-b^2+\sqrt{(z^2-b^2)(1-b^2)}}
\right.\nonumber\\
&&\left.
+\sqrt{\frac{z^2-b^2}{1-b^2}}\log(1-b^2)
-\mu\left(
z-\sqrt{\frac{z^2-b^2}{1-b^2}}
\right)
\right].
\eeqa
Since \beqa
\mu=-\log(1-b^2),
\eeqa
$y(z)$ is simply given by 
\beqa
y(z)&=&W'(z) -2\mu\omega(z)\nonumber\\
&=&-\frac2{1-z^2}\left(
\log\frac{-\sqrt{z^2-b^2}+\sqrt{1-b^2}}{\sqrt{z^2-b^2}+\sqrt{1-b^2}}
\right).
\eeqa
Going back to the original $u$ plane by the relation
\beqa
z&=&\tanh\frac u2~=~\frac{e^u-1}{e^u+1}, 
\eeqa
the differential $y(z)dz$ becomes
\beqa
y(z)dz &=&
2 du\left(
\log\frac{1+e^{-u}+\sqrt{(1+e^{-u})^2-4 e^{\mu-u}}}2
+\frac{u-\mu}2
\right).
\eeqa
Thus one recovers the equation given in \cite{AKMV} 
describing the mirror of a resolved conifold \cite{HV}, 
which can be made identical to the equation of a deformed conifold 
by a change of variables.  

\subsection{The Special Geometry Relation and Free Energies}
The special geometry relation for this matrix model can be similarly 
obtained, although the branch cuts of $W'(z)$, which run from $-1$ to 
$-\infty$ and from $+1$ to $+\infty$, add an extra complication; one must 
specify the relative location of the branch cut of $\log|\lambda|$ 
(taking $\lambda_{01}\equiv 0$) in (\ref{F0multicut}) and define the contour 
$\hat B_1$ accordingly.
Then one differentiates $F_0$ by $\mu$ and employ an identity 
similarly derived by using Riemann's bilinear relation. 
Although tedious, the calculation can be done similarly as before and 
we only describe the result:
\beqa
\frac{\partial F_0}{\partial \mu}&=&
 \lim_{\epsilon\rightarrow 0}
\left[
\int_b^{1-\epsilon} dz y(z)
-W(1-\epsilon)
\right],\label{dF0dmuCS}
\\
\frac{\partial y(z)}{\partial \mu}dz
&=&\omega_{1-\widetilde{1}}(z) +\omega_{(-1)-\widetilde{(-1)}}(z),
\label{dydmuCS}
\eeqa
where, as before, 
$\omega_{1-\widetilde{1}}(z)$ ($\omega_{(-1)-\widetilde{(-1)}}(z)$) 
is a zero A-period abelian differential of the third kind having two simple poles, 
one at $z=1$ ($z=-1$) with residue $1$ and the other at $z=\widetilde1$ 
($z=\widetilde{-1}$) with residue $-1$.
Differentiating (\ref{dF0dmuCS}) by $\mu$ again and using (\ref{dydmuCS}),
we find
\beqa
\frac{\partial^2 F_0}{\partial\mu^2}
&=&
\lim_{\epsilon\rightarrow 0}
\left[
\int_b^{1-\epsilon} dz\sqrt{\frac{1-b^2}{z^2-b^2}}
\left(
\frac1{z-1}-\frac1{z+1}
\right)
-\log 2\epsilon
\right]\nonumber\\
&=&\log\frac{b^2}{4(1-b^2)}.
\eeqa
Integrating over $\mu$ yields
\beqa
F_0&=&
-\sum_{n=1}^\infty\frac{e^{-n\mu}}{n^3}
+\frac{\mu^3}6
-\frac{\mu^2}2\log 4
+\mbox{const}\times\mu~+\mbox{const}.
\eeqa
$F_{0\mbox{\scriptsize CS}}=F_0-\frac{\mu^3}{12}$ agrees with the known 
result \cite{Witten,GopakumarVafa}.

The genus-1 free energy may also be computed by the formula 
of Ambj\o rn et. al. \cite{ACKM} again:
\beqa
F_1&=&-\frac1{24}(\log M_1 +\log J_1 +4\log 2b).
\eeqa
After some calculations we find 
\beqa
M_1~=~J_1&=&\frac4{\mu\sqrt{1-b^2}^3},
\eeqa
and hence obtain
\beqa
F_1&=&-\frac1{12}\left(
\frac32 \mu -\log\mu
+\log(1-e^{-\mu})
+4\log2
\right),
\eeqa
which correctly reproduces the known genus-1  Chern-Simons 
free energy $F_{1\mbox{\scriptsize CS}}=F_1+\frac\mu{12}$
including the $\log\mu$ term \cite{Witten,GopakumarVafa}. 
\section{Conclusions}
In this paper we have shown that a hermitian 1-matrix model with 
the potential 
$W(\Phi)$
and a unitary 1-matrix model with the potential 
$W_U(U)$
are nothing but different descriptions of the same model
if they are related with each other by the equation (\ref{W}).
The extra logarithmic term on the hermitian side has been further   
interpreted as an effect of the coupling to $2N$ vectors. 
This duality is particularly useful to compute lower genus free 
energies 
in unitary as well as 
other matrix models with the unitary matrix Haar measure, and we have 
demonstrated the use of it in various examples. It allowed us not only to
rederive the known facts by simply using the familiar hermitian matrix model 
technologies, but also to establish some new results on matrix models and 
gauge theory. We have shown that 
the unitary matrix formulation of the ${\cal N}=2$ pure 
$SU(2)$ gauge theory correctly reproduces the genus-1 amplitude, and 
also given the rigorous forms of the special geometry relations 
in unitary as well as the Chern-Simons matrix models. 

This duality is not only of practical but also of conceptual 
importance; it has revealed that unitary and hermitian 1-matrix models 
have no essential difference and can be mapped to each other via 
the transformation (\ref{W}). An immediate consequence of this fact 
is that the partition function of a unitary 1-matrix model is also a $\tau$
function of the KP hierarchy. It also provides an explanation of  
why the critical behaviors of both unitary and (even-potential) 
hermitian 1-matrix models are described by the same (Painlev\'e II) equation 
when two clusters of eigenvalues join: this is accounted for by the universality 
with respect to even perturbations (e.g. \cite{HMPN}) in hermitian 1-matrix models. 

It is amusing that 
while the topological B-model on a deformed conifold
is known to be equivalent to $c=1$ strings at the self-dual 
radius \cite{GhoshalVafa} and realized as the Penner model \cite{DistlerVafa},
the hermitian matrix model obtained after applying the 
duality to the Chern-Simons matrix model
may be regarded as a ``two-flavor'' extension of the 
(generalized) Penner model.

It would be interesting to apply this duality to the matrix models 
proposed to describe some five-dimensional gauge theories 
and compare the lower genus amplitudes with Nekrasov's formula 
\cite{Nekrasov}.

\section*{Acknowledgments}
I would like to thank T.~Kimura for discussions and C.~V.~Johnson for 
correspondence. This work was supported in part by Grant-in-Aid
for Scientific Research (C)(2) \#16540273 from
The Ministry of Education, Culture, Sports, Science
and Technology.

\section*{Appendix A~ The Multi-cut Free Energy}
\setcounter{equation}{0}
\def\theequation{A.\arabic{equation}}
\def\thesection{A}
In this Appendix we give a derivation of the formula (\ref{F0multicut}) for 
the genus-0 multi-cut free energy in hermitian 1-matrix models.
We start from the well-known expression \cite{BIPZ}
\beqa
\frac{F_0}{\mu^2}&=&-\frac1\mu \int_{-\infty}^\infty d\lambda
\rho(\lambda) W(\lambda)
+ \int_{-\infty}^\infty d\lambda
\int_{-\infty}^\infty d\lambda'\rho(\lambda)\rho(\lambda')\log|\lambda-\lambda'|,
\label{F0BIPZ}
\eeqa
which is valid for any number of the cuts $n$. 
It is convenient to define 
\beqa
\sigma(\lambda)&\equiv&\int_{-\infty}^\lambda
 d\lambda' \rho(\lambda').
 \eeqa
 Due to (\ref{muj}) we have 
 \beqa
 \sigma(a_1)&=&0,\nonumber\\
 \sigma(a_2)&=&\sigma(a_3)~=~\frac{\mu_1}\mu,\nonumber\\
 \cdots,&&\nonumber\\
 \sigma(a_{2j})&=&\sigma(a_{2j+1})~=~\sum_{k=1}^j\frac{\mu_k}\mu,\nonumber\\
 \cdots,&&\nonumber\\
 \sigma(a_{2n})&=&1.
 \eeqa
 Using $\sigma(\lambda)$, the second term of (\ref{F0BIPZ}) can be 
 integrated by parts to give
\beqa
\mbox{The second term}&=&
\int_{-\infty}^\infty d\lambda \rho(\lambda)
\sum_{j=1}^n\int_{a_{2j-1}}^{a_{2j}} 
d\lambda'\rho(\lambda')\log|\lambda-\lambda'|
\nonumber\\
&=&
\int_{-\infty}^\infty d\lambda \rho(\lambda)
\sum_{j=1}^n
\left(
\left[\rule{0mm}{5mm}
\sigma(\lambda')\log|\lambda-\lambda'|
\right]_{a_{2j-1}}^{a_{2j}}
+\int_{a_{2j-1}}^{a_{2j}}
d\lambda'\frac{\sigma(\lambda')}{\lambda-\lambda'}
\right)
\nonumber\\
&=&
\int_{-\infty}^\infty d\lambda \rho(\lambda)
\sum_{j=1}^n
\left(
\frac1\mu\sum_{k=1}^j \mu_k \log|\lambda-a_{2j}|
-\frac1\mu\sum_{k=1}^{j-1} \mu_k \log|\lambda-a_{2j-1}|
\right)\nonumber\\
&&
-\sum_{j=1}^n\int_{a_{2j-1}}^{a_{2j}}
d\lambda \sigma(\lambda)\frac1{2\mu} W'(\lambda),
\eeqa
where in the last line we have used the large $N$ saddle point 
equation 
\beqa
-\frac1\mu W'(\lambda)+2\int_{-\infty}^\infty
\frac{\rho(\lambda')}{\lambda-\lambda'}&=&0.
\label{saddlept}
\eeqa
The last term is integrated by parts again to yield
\beqa
-\sum_{j=1}^n\int_{a_{2j-1}}^{a_{2j}}
d\lambda \sigma(\lambda)\frac1{2\mu} W'(\lambda)
&=&
\frac1{2\mu}\int_{-\infty}^\infty d\lambda \rho(\lambda)V(\lambda)
-\frac1{2\mu}\left(
\frac{\mu_1}\mu(W(a_2)-W(a_3))+\cdots
\right.\nonumber\\  &&\left.~~~~~~
+\frac{\sum_{k=1}^{n-1}\mu_k}\mu(W(a_{2n-2})-W(a_{2n-1}))
+W(a_{2n})
\right).
\eeqa
Therefore we get
\beqa
\frac{F_0}{\mu^2}&=&
-\frac1{2\mu}\int_{-\infty}^\infty d\lambda \rho(\lambda)W(\lambda)\nonumber\\
&&+\frac{\mu_1}\mu
\left(
-\frac1{2\mu}(W(a_2)-W(a_3))+\int_{-\infty}^\infty \rho(\lambda)
\log\left|
\frac{\lambda-a_2}{\lambda-a_3}
\right|
\right)
\nonumber\\
&&+\cdots\nonumber\\
&&+\frac{\sum_{k=1}^j\mu_k}\mu
\left(
-\frac1{2\mu}(W(a_{2j})-W(a_{2j+1}))+\int_{-\infty}^\infty \rho(\lambda)
\log\left|
\frac{\lambda-a_{2j}}{\lambda-a_{2j+1}}
\right|
\right)
\nonumber\\
&&+\cdots\nonumber\\
&&+\left(
-\frac1{2\mu}W(a_{2n})+\int_{-\infty}^\infty \rho(\lambda)
\log\left|
\lambda-a_{2n}
\right|
\right).
\eeqa
Here the saddle point equation (\ref{saddlept}) can be used to replace 
$a_{2k-1}$, $a_{2k}$ with an arbitrary $\lambda_{0k}$ on the $k$th cut, 
for all $k=1,\ldots,n$. With this replacement we end up with the formula
(\ref{F0multicut}).

\section*{Appendix B~ A Derivation of Eq.(\ref{Riemann'srelation})}
\setcounter{equation}{0}
\def\theequation{B.\arabic{equation}}
\def\thesection{B}
Let $\Sigma$ be a hyper-elliptic Riemann surface defined by  
\beqa
Y^2&=&\prod_{k=1}^{2n}(z-a_k).
\eeqa
The first step in using Riemann's bilinear relation is to cut out $\Sigma$ 
along the cycle
\beqa
A_{n-1}B_{n-1}A_{n-1}^{-1}B_{n-1}^{-1}
\cdots
A_{1}B_{1}A_{1}^{-1}B_{1}^{-1}
~
\hat B_n P_-^{-1} A_\infty P_+^{-1}
\hat B_n^{-1} P_{\tilde-}^{-1} A_{\tilde \infty} P_{\tilde+}^{-1}
\eeqa
to get a simply-connected polygonal region, 
which we denote by $\Sigma'$. For the definitions of the cycles see 
Figures 1 and 2. 
\begin{figure}
\centering
\resizebox{0.65\textwidth}{!}{%
  \rotatebox{0}{\includegraphics{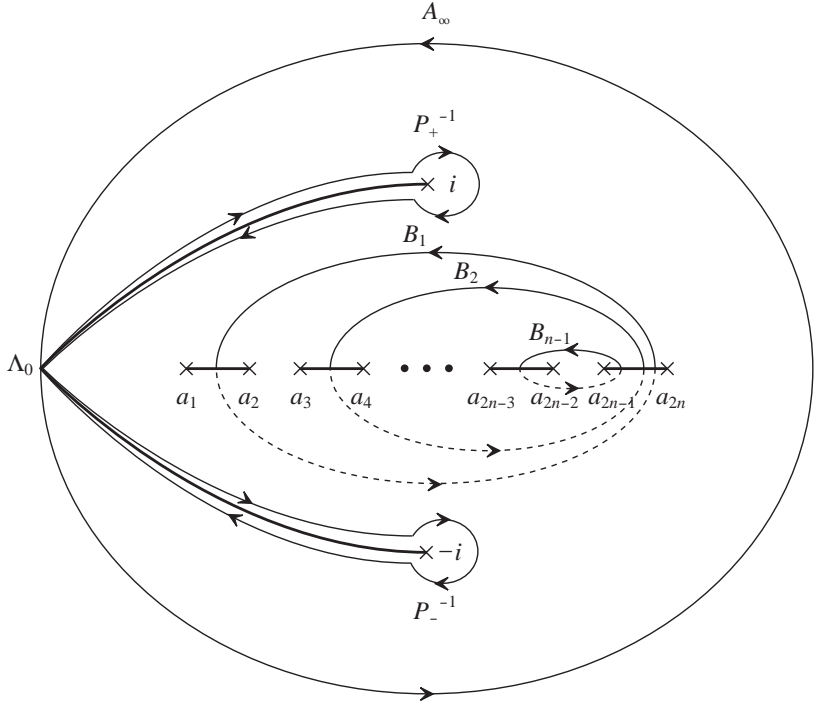}}}
\caption{More definitions of the contours. 
The bold lines from $\pm i$ to $\Lambda_0$ denote the branch cut of 
$\log(z\mp i)$. $A_{\tilde\infty}$, $P_{\tilde\pm}^{-1}$ are the same as those 
without tildes but lying on the second sheet. }
\end{figure}
Next we consider 
\beqa
\Omega_i(z)&\equiv&\int_{z_0}^z \omega_i~~~(i=1,\ldots,n-1)
\eeqa
for some reference point $z_0$ on $\Sigma'$, then $\Omega_i(z)$'s 
are single-valued, everywhere holomorphic functions {\it inside} $\Sigma'$. 
Then for each $i$ Riemann's bilinear relation can be obtained by computing
\beqa
\int_{\partial\Sigma'} dz y(z)\Omega_i(z)
\eeqa
in two ways, by the residue computation inside $\Sigma'$, and by 
evaluating the integral along the boundary of  $\Sigma'$.
The total residue is trivially zero. On the other hand, one can evaluate 
the integral on the edges pairwise as
\beqa
\left(\oint_{A_j} +\oint_{A_j^{-1}}\right) dz y(z)\Omega_i(z)
&=&4\pi i\mu_j\oint_{B_j}\omega_i,\nonumber\\
\left(\oint_{B_j} +\oint_{B_j^{-1}}\right) dz y(z)\Omega_i(z)
&=&\delta_{ij}\oint_{B_j}y(z),\nonumber\\
\left(\int_{\hat B_n} +\int_{\hat B_n^{-1}}\right) dz y(z)\Omega_i(z)
&=&0,\nonumber\\
\Lim\left(\int_{\hat P_-^{-1}} +\int_{\hat P_+^{-1}}\right) dz y(z)\Omega_i(z)
&=&-\sum_{j=1}^n \oint_{A_j} W(z)\omega_i(z) -4\pi i\mu\Omega_i(-\infty),
\nonumber\\
\Lim\left(\int_{\hat P_{\tilde-}^{-1}} +\int_{\hat P_{\tilde+}^{-1}}\right) dz y(z)\Omega_i(z)
&=&-\sum_{j=1}^n \oint_{A_j} W(z)\omega_i(z) +4\pi i\mu\Omega_i(\widetilde{-\infty}),
\eeqa
and
\beqa
\oint_{A_\infty}  dz y(z)\Omega_i(z)
&=&\oint_{A_{\tilde\infty}}  dz y(z)\Omega_i(z)
~=~0.
\eeqa 
Stokes' theorem implies that the sum of all these integrals is zero.
Noting that
\beqa
\Omega_i(\infty)-\Omega_i(\widetilde{-\infty})&=&\Lim\int_{\hat B_n} \omega_i,
\eeqa
 we have
\beqa
-4\pi i\sum_{j=1}^n \mu_j \Lim\int_{\hat B_j} \omega_i
+\oint_{B_i} dz y(z) -2\sum_{j=1}^n \oint_{A_j} W(z) \omega_i(z)&=&0
\label{(1)}
\eeqa
for $i=1,\ldots,n-1$.

To derive a relation involving the period of $\hat\omega_n$ we also 
consider 
\beqa
\int_{\partial\Sigma'} dz y(z)\Omega_n(z),
\label{intyOmegan}
\eeqa
where
\beqa
\Omega_n(z)&\equiv&\int_{z_0}^z \hat\omega_n
\eeqa
is also a single-valued holomorphic function inside $\Sigma'$. 
Taking special care about the singular behavior of $y(z)\Omega_n(z)$ 
near $z=\pm i$, one may similarly compute (\ref{intyOmegan}) in two ways 
to obtain
\beqa
&&\hskip -10mm 
-\frac12 \sum_{j=1}^n \oint_{A_j} W(z) \hat\omega_n(z)
+\Lim\left(
-\pi i\sum_{j=1}^n\mu_j \int_{\hat B_j}\hat\omega_n
\right)\nonumber\\
&&=\Lim\left[
\frac14 \int_{\hat B_n} dz y(z)
\right.\nonumber\\
&&\left.
\hskip 18mm+\frac1{8\pi i}\left(
\int_{P_-^{-1}}\!\!\!+\int_{P_+^{-1}}
\right)dz (y(z)-W'(z)) \log(z^2 +1)
-\frac12 W(\Lambda_0)
\right].\label{(2)}
\eeqa
One can then use (\ref{(1)}) in this equation to see also that 
\beqa
&&\hskip -10mm 
-\frac12 \sum_{j=1}^n \oint_{A_j} W(z) \hat\omega_i(z)
+\Lim\left(
-\pi i\sum_{j=1}^n\mu_j \int_{\hat B_j}\hat\omega_i
\right)\nonumber\\
&&=\Lim\left[
\frac14 \int_{\hat B_i} dz y(z)
\right.\nonumber\\
&&\left.
\hskip 18mm+\frac1{8\pi i}\left(
\int_{P_-^{-1}}\!\!\!+\int_{P_+^{-1}}
\right)dz (y(z)-W'(z)) \log(z^2 +1)
-\frac12 W(\Lambda_0)
\right].
\eeqa
for $i=1,\ldots,n-1$, and hence, due to (\ref{(2)}), $i=1,\ldots,n$.
Since the last line is equal to
\beqa
\lim_{\epsilon\rightarrow 0}\left[\frac14\Lim
\left(\int_{\Lambda_0}^{i-\epsilon}
+\int_{\Lambda_0}^{-i-\epsilon}\right)
dz y(z) -W(i-\epsilon)-W(-i-\epsilon)
\right],
\eeqa
we obtain the desired equation  (\ref{Riemann'srelation}).
%
%

\end{document}